\documentclass[showpacs,twocolumn,prl,aps]{revtex4}
\usepackage{graphicx}

\begin{document}

\title[Short title for running header]{Metal-insulator transition in quarter-filled Hubbard model on triangular lattice and its implication for the
physics of $Na_{0.5}CoO_{2}$}
\author{Tao Li}
\affiliation{Department of Physics, Renmin University of China,
Beijing 100872, P.R.China}
\date{\today}

\begin{abstract}
The metal-insulator transition of the quarter-filled Hubbard model
on triangular lattice is studied at the mean field level. We find a
quasi-one dimensional metallic state with a collinear magnetic order
competes closely with an insulating state with a non-coplanar
magnetic order for both signs of the hopping integral $t$. In the
strong correlation regime($U/|t|\gg1$), it is found that the
metal-insulator transition of the system occurs in a two-step
manner. The quasi-one dimensional metallic state with collinear
magnetic order is found to be stable in an intermediate temperature
region between the paramagnetic metallic phase and the non-coplanar
insulating phase. Possible relevance of these results to the physics
of metal-insulator transition in $Na_{0.5}CoO_{2}$ is discussed.
\end{abstract}
\maketitle

$Na_{x}CoO_{2}$ is a remarkable transition metal oxide system in
which strong electronic correlation and geometric frustration
interplay with each other. Unconventional superconductivity, novel
magnetic and charge ordered states and state with peculiar
Curie-Weiss metal behavior are observed in this system at different
carrier
concentrations\cite{nco1,nco2,nco3,nco4,nco5,nco6,nco7,nco8}. To
date, most of these observations remains not well understood.

The quarter-filled system, namely $Na_{0.5}CoO_{2}$ is especially
interesting\cite{nco3,nco6,nco8}. It divides the phase diagram of
the $Na_{x}CoO_{2}$ system into two parts of qualitatively different
nature. For $x<0.5$, the system exhibits conventional metallic
behavior in the normal state and with water intercalation a
superconducting state with unknown paring symmetry in a small doping
region around $x=0.3$\cite{nco1}. While fore $x>0.5$, the system
exhibits Curie-Weiss metal behavior at high temperature and a weak
static magnetic order at low temperature with ferromagnetic
intra-layer spin correlation\cite{nco4,nco5,nco7}. Right at $x=0.5$,
the observations are even more striking. The system undergoes two
phase transitions at 88K and 53K\cite{nco3,nco6}. At the 88K
transition, the system develops a magnetic order with a ordering
wave vector
$\mathrm{\mathbf{q}}=\frac{1}{2}(\mathrm{\vec{\mathbf{b}}}_{1}+\mathrm{\vec{\mathbf{b}}}_{2})$\cite{qunit}.
However, the longitudinal resistivity of the system seems to be
totally ignorant of the transition. While at the 53K transition, the
system enters the true insulating state as signaled by the rapid
increase of the longitudinal resistivity below this temperature.
However, both the peak position and the intensity of the magnetic
Bragg peak developed at 88K seems to be ignorant of the transition
at 53K. The observation of two transitions and the apparent mutual
independence of the magnetic order and the charge transport have
arose many interest in the field. However, this is still not the
whole story. In fact, it is found that Hall resistivity begins to
drop abruptly around the 88K transition which indicates that it is
not a transition in the spin sector only\cite{nco3}.

No coherent picture exists yet on the low temperature phase
transitions of the $Na_{0.5}CoO_{2}$ system. It is generally
anticipated that the ordering of the Na ions outside the $CoO_{2}$
plane may plays an important role in the low temperature physics of
this system\cite{nco3,nco6,nco10,nco12}. Charge ordering(induced by
the Sodium order) in the $CoO_{2}$ plane is invoked in many works to
explain the multiple phase transitions. In this paper, we suggest a
possible picture for the multiple phase transitions of the
$Na_{0.5}CoO_{2}$ system as an \textit{intrinsic} property of a
quarter-filled strongly correlated system on the triangular lattice.
We backup our reasoning with a mean field study of the
metal-insulator transition of the quarter-filled Hubbard model on
triangular lattice. We find this seemingly oversimplified model does
exhibit two successive transitions at quarter filling in its way
toward the low temperature insulating state. In our picture, the
magnetic transition at 88K transform the system into an quasi-one
dimensional metallic state through a \textit{spin ordering induced
dimensional reduction} process. This quasi-one dimensional metal can
then transform into the true insulating state at 53K in various
possible ways. We thus predict that the intermediate phase between
53K and 88K to be a state with one dimensional transport property.

\begin{figure}[h!]
\caption{(a)The magnetic structure of the proposed state for the
intermediate temperature region between 53K and 88K. Note the state
has no charge order and all lattice sites are still uniformly
quarter-filled. (b)The insulating state with a four sublattice and
non-coplanar magnetic structure predicted by the mean field theory
of the quarter-filled Hubbard model on triangular lattice. The
magnetic moments on the four sublattices point to the four corners
of a perfect tetrahedron as shown in the inset.} \label{fig1}
\end{figure}

The state we propose for the intermediate temperature region between
53K and 88K is shown schematically in Figure 1a. In this state, the
system self-organizes into a series of chains with ferromagnetic
order on them. Neighboring chains are antiferromagnetically
correlated with each other. Unlike other proposals, there is no
charge ordering in this state and all lattice sites are still
quarter-filled uniformly and thus the ferromagnetic chains are
metallic. It is interesting to note that this metallicity is of one
dimensional nature as any inter-chain electron hopping must cross
the spin gap. Thus the spin ordering in such a particular pattern
renders the system a one dimensional metal and so we dub it as a
\textit{spin ordering induced dimensional reduction} process.
Similar dimensional reduction process also occurs in other condensed
matter system with multiple low energy degree of freedoms. The most
well known example is the orbital ordering induced dimensional
reduction in ruthenium oxides\cite{nco14}.

The magnetic structure of the state shown in Fig.1a is given by
$<\mathrm{\vec{S}}_{i}>=\vec{\mathbf{m}}\exp(i\mathbf{Q}_{3}\cdot\mathbf{R}_{i})$
with
$\mathrm{\mathbf{Q}}_{3}=\frac{1}{2}(\vec{\mathbf{b}}_{1}+\vec{\mathbf{b}}_{2})$.
The states with $\mathbf{Q}_{1}=\frac{1}{2}\vec{\mathbf{b}}_{1}$ and
$\mathbf{Q}_{2}=\frac{1}{2}\vec{\mathbf{b}}_{2}$ are degenerate with
it by lattice symmetry. All the three states are consistent with the
neutron scattering measurement and the real system will contain
domains of all three kinds. Since the ferromagnetic chains are still
metallic, the longitudinal resistivity of the system is not expected
to change dramatically across the 88K transition. However, the
transverse transport is totally different. As the system becomess
quasi-one dimensional in charge transport state, we expect the Hall
resistivity to drop abruptly at the 88K transition, in accordance
with the experimental observation. Thus the state shown in Fig.1a
has the correct characteristics to account for the phenomenology of
the intermediate phase between 53K and 88K.

With further lowering of temperature, the quasi-one dimensional
intermediate state will eventually become a true insulator. It is
important to note that the intermediate state has an enlarged unit
cell with two lattice sites. Thus the system is half-filled below
the 88K transition. To arrive at the true insulating state, the
system should lower its symmetry further to have a unit cell with at
least four sites. Through mean field calculation on the Hubbard
model on triangular lattice, we find a state with a four sublattice
structure and non-coplanar spin configuration (see Fig.1b) that meet
this requirement. This state has no charge order either and the
magnetic structure can be written as
$<\mathrm{\vec{S}}_{i}>=\vec{\mathbf{m}}_{1}\exp(i\mathbf{Q}_{1}\cdot\mathbf{R}_{i})
+\vec{\mathbf{m}}_{2}\exp(i\mathbf{Q}_{2}\cdot\mathbf{R}_{i})+\vec{\mathbf{m}}_{3}\exp(i\mathbf{Q}_{3}\cdot\mathbf{R}_{i})$.
Here the three vectors
$\vec{\mathbf{m}}_{1},\vec{\mathbf{m}}_{2},\vec{\mathbf{m}}_{3}$ are
of same length and are mutually orthogonal. The magnetic moments in
the four sublattices points to the four corners of a perfect
tetrahedron. This state has the same set of magnetic Bragg peaks as
the collinear state shown in Fig.1a if the three kinds of domains of
the collinear state appear with the same probability. The mean field
theory presented in this paper also indicates that the transition
between collinear state and the four sublattice insulating state
leaves the intensity of the magnetic Bragg peak almost intact. This
is in accordance with the observation at the 53K transition.

It should be noted that both kinds of order shown in Fig.1 have
already been predicted in a mean field study of the quarter-filled
Hubbard model on triangular lattice with a \textit{negative} hopping
integral\cite{nco12}. In that work, both kinds of order develop as a
result of the nesting property of the Fermi surface and exist for
any non-zero interaction. However, the Fermi surface of the
$Na_{0.5}CoO_{2}$ is almost featureless. As will be shown below, the
$Na_{0.5}CoO_{2}$ system should be described as a quarter-filled
system with a \textit{positive} hopping integral. Since the
triangular lattice has no particle-hole symmetry, any magnetic order
can develop only for interaction greater than certain critical
value. Indeed, we find the magnetic structures shown in Fig.1 exist
only in the strong correlation regime for $Na_{0.5}CoO_{2}$.

We now present the results of the mean field study of the
metal-insulator transition in the quarter-filled Hubbard model on
triangular lattice. The Hamiltonian of the model reads
\begin{equation}
\mathrm{H}=-t\sum_{<i,j>,\sigma}(c^{\dagger}_{i,\sigma}c_{j,\sigma}+h.c.)+U\sum_{i}n_{i,\uparrow}n_{i,\downarrow}.
\end{equation}
It is generally believed that the $Na_{x}CoO_{2}$ system resides in
the strongly correlated regime with $\frac{U}{t}>>1$\cite{nco9}.
Since the triangular lattice has no particle-hole symmetry, the sign
of the hopping integral $t$ is important. For $Na_{x}CoO_{2}$
system, it is negative as inferred from ARPES
measurement\cite{nco11}. In $Na_{0.5}CoO_{2}$, each $Co$ site hosts
$1.5$ electrons on average is thus a $\frac{3}{4}$-filled system
with a negative hopping integral. After a particle-hole
transformation, it is equivalent to a quarter-filled system with a
positive hopping integral. For purpose of comparison, we also
consider the case of quarter-filled system with a negative hopping
integral which is studied already in Ref 12.

From the weak coupling point of view, a quarter-filled system on
triangular lattice with a positive hopping integral has no
speciality at all as compared to a general incommensurate filling.
The Fermi surface has a almost perfect rounded shape. Thus, it is
hard to guess what would happen in the strong correlation regime
from weak coupling analysis. To the contrary, a quarter-filled
system on triangular lattice with a negative hopping integral is
special in the sense that the chemical potential rests just on the
Van Hove singularity of the density of state($\mu=-2|t|$ at zero
temperature). The Fermi surface is nested with nesting vectors
$\mathbf{Q}_{1},\mathbf{Q}_{2}$, and $\mathbf{Q}_{3}$. As the result
of the nesting property of the Fermi surface, the system develops
magnetic order at an arbitrarily small $U$. In Ref. 12, the collinear
state shown in Fig. 1a and the non-coplanar state shown in Fig.1.b
are found to be the most favorable choice for the magnetic
structure. Among the two, the non-coplanar state is found to be
slightly more favorable.

To have an idea on what would happen in the strong correlation
regime for the quarter-filled Hubbard system with a positive hopping
integral, we have conducted a unrestricted mean field search at zero
temperature on a $12\times12$ lattice. The search is done in the
mean field space with the local density $n_{i}$ and the three
components of local spin density $\vec{S}_{i}$ as variational
parameters. We have used both the conjugate gradient method and the
simulated annealing method to perform the search and used more than
100 random initial configurations for the local density and local
spin density. We find the search always converge to the two
configurations shown in Fig.1(up to global spin rotations and point
group operations) for large enough value of $\frac{U}{t}$. Thus,
although the quarter-filled system with positive hopping integral
has no nesting property, it has the same magnetic ordering pattern
in the strong correlation regime as the negative hopping system.

The mean field equations for the collinear state is given by
\begin{equation}
1=\frac{2U}{N}\sum_{\mathbf{k}}\left[\frac{(\xi_{\mathbf{k}}-E^{+}_{\mathbf{k}})f(E^{+}_{\mathbf{k}})}{(\xi_{\mathbf{k}}-E^{+}_{\mathbf{k}})^{2}+(Um)^{2}}
+\frac{(\xi_{\mathbf{k}}-E^{-}_{\mathbf{k}})f(E^{-}_{\mathbf{k}})}{(\xi_{\mathbf{k}}-E^{-}_{\mathbf{k}})^{2}+(Um)^{2}}\right],
\end{equation}
in which the sum is limited in the magnetic Brillouin zone and
$\xi_{\mathbf{k}}=-2t(\cos\mathrm{k}_{x}+\cos\mathrm{k}_{y}+\cos(\mathrm{k}_{x}+\mathrm{k}_{y}))-\mu$
is the dispersion on the triangular lattice. Here we have used the
convention for the momentum that
$\mathbf{k}=\mathrm{k}_{x}\mathbf{\vec{b}}_{1}+\mathrm{k}_{x}\mathbf{\vec{b}}_{2}$.
In this convention, the first Brillouin zone of the triangular
lattice is given by
$(\mathrm{k}_{x},\mathrm{k}_{y})\in[-\pi,\pi]\otimes[-\pi,\pi]$ and
the magnetic Brillouin zone of the collinear state is given by
$[0,\pi]\otimes[-\pi,\pi]$.
$E_{\mathbf{k}}^{\pm}=\frac{\xi_{\mathbf{k}}+\xi_{\mathbf{k+Q_{3}}}\pm\sqrt{(\xi_{\mathbf{k}}-\xi_{\mathbf{k+Q_{3}}})^{2}+(Um)^{2}}}{2}$
is quasiparticle energy in the magnetic ordered state.

In the non-coplanar insulating state, the magnetic Brillouin zone
reduces further into the region given by $[0,\pi]\otimes[0,\pi]$.
The mean field equation for the non-coplanar state is given by
\begin{equation}
m=\frac{1}{N}\sum_{\mathbf{k},n}W^{n}_{\mathbf{k}}f(E^{n}_{\mathbf{k}}),
\end{equation}
in which
$W^{n}_{\mathrm{\mathbf{k}}}=\frac{1}{2\sqrt{3}}\sum_{\alpha,\beta}\phi_{\alpha}^{n*}(\mathrm{\mathbf{k}})\mathcal{M}_{\alpha,\beta}\phi_{\beta}^{n}(\mathrm{\mathbf{k}})$
, $\alpha,\beta,n=1\cdots8$ and the sum is limited in the magnetic
Brillouin zone. $\phi_{\alpha}^{n}(\mathrm{\mathbf{k}})$ is the
$\alpha-th$ component of the eigenvector for the mean field
Hamiltonian with eigenvalue $E^{n}_{\mathrm{\mathbf{k}}}$. The mean
field Hamiltonian matrix is of the following form
\begin{equation}
\mathrm{H_{MF}} = \left( {\begin{array}{*{20}c}
   \xi_{\mathbf{k}}\mathrm{I} & \mathrm{M}_{1}  & \mathrm{M}_{2}^{*} & \mathrm{M}_{2}\\
   \mathrm{M}_{1} & \xi_{\mathbf{k}+\mathbf{Q}_{1}}\mathrm{I}  & \mathrm{M}_{2} &\mathrm{M}_{2}^{*}\\
   \mathrm{M}_{2}^{*} & \mathrm{M}_{2}  & \xi_{\mathbf{k}+\mathbf{Q}_{2}}\mathrm{I} & \mathrm{M}_{1}\\
   \mathrm{M}_{2} & \mathrm{M}_{2}^{*}  & \mathrm{M}_{1} & \xi_{\mathbf{k}+\mathbf{Q}_{3}}\mathrm{I}\\
 \end{array} } \right),
\end{equation}
here $\mathrm{I}$ is a $2\times2$ identity matrix, $
\mathrm{M}_{1}=-\frac{Um}{3}\left( {\begin{array}{cc}
   1 & \sqrt{2}\\
   \sqrt{2} & -1\\
 \end{array} } \right)$, $\mathrm{M}_{2}=-\frac{Um}{3}\left( {\begin{array}{cc}
   1 & \frac{-1+i\sqrt{3}}{\sqrt{2}}\\
   \frac{-1-i\sqrt{3}}{\sqrt{2}} & -1\\
 \end{array} } \right)$. The matrix $\mathcal{M}$ is given by
\begin{equation}
\mathcal{M} = -\frac{3}{Um}\left( {\begin{array}{*{20}c}
   0 & \mathrm{M}_{1}  & \mathrm{M}_{2}^{*} & \mathrm{M}_{2}\\
   \mathrm{M}_{1} & 0  & \mathrm{M}_{2} &\mathrm{M}_{2}^{*}\\
   \mathrm{M}_{2}^{*} & \mathrm{M}_{2}  & 0 & \mathrm{M}_{1}\\
   \mathrm{M}_{2} & \mathrm{M}_{2}^{*}  & \mathrm{M}_{1} & 0\\
 \end{array} } \right) \nonumber.
\end{equation}

\begin{figure}[h!]
\caption{The order parameters of magnetic ordered states at zero
temperature as functions of the coupling strength $U/|t|$ for both
signs of hopping integral. (a)$t>0$, (b)$t<0$. } \label{fig2}
\end{figure}

Fig.2 shows the order parameter $m$ as a function of $U/|t|$ at zero
temperature for both signs of the hopping integral. For negative $t$
, the magnetic order exist for all non-zero $U$ as a result of the
nesting property of the system. While for positive $t$, the magnetic
order exists only for $U/|t|>U_{c}/|t|=7.2$. For both sign of $t$,
the non-coplanar state is slightly more stable than the collinear
state at zero temperature.

Fig.3 shows the mean field phase diagram at finite temperature for
both signs of the hopping integral. At finite temperature, the
gapless Fermion excitation in the collinear state will contribute
more entropy than the gapped Fermion excitation in the non-coplanar
insulating state. We find for sufficiently large $U/|t|$, the
collinear state will become more stable than the non-coplanar state
above a critical temperature at which a first order transition
between the two occurs. In the phase diagram, the first order
transition line between the collinear state and the non-coplanar
state end at a critical end point on the magnetic-paramagnetic
transition line. For positive $t$, the critical end point locates at
$U/|t|=7.9$, while for negative $t$, it locates at a smaller value
of $U/|t|=5.5$. Thus for both sign of the hopping integral, the
collinear state is stable only in the strong correlation regime.

It is interesting to note that the transition between the collinear
state and non-coplanar state, though is of first order in nature,
has almost no observable effect on the peak position and intensity
of the magnetic Bragg peak. As we have explained above, the
non-coplanar state has the same sets of magnetic Bragg peak as the
collinear state, if the three domains of the latter as mixed with
equal probability. In Fig.4, we plot the change of the magnitude of
the order parameter at the collinear - non-coplanar phase transition
as a function of $U/|t|$. We find the change to be always positive
but its absolute value is always less than $1.5\%$ of the ordered
moment for all value of $U/|t|$ and for both signs of $t$. Thus we
conclude that the neutron scattering experiments are not sensitive
to the collinear-non-coplanar transition.

However, as the system opens a full gap abruptly on the
quasiparticle spectrum at the collinear-non-coplanar transition, the
longitudinal resistivity is expected to increase dramatically. Below
the transition point, the system will enter the true insulating
state. Thus in this system, the metal-insulator transition occurs in
a two-step manner. The quarter-filled system first breaks the
lattice symmetry by a two-fold enlargement of the unit cell and
enters a half-filled quasi-one dimensional metallic state at the
paramagnetic-magnetic transition and then reduces its symmetry
further by again a two-fold enlargement of the unit cell at the
collinear-non-coplanar transition and enters a integer-filled
insulating state. It is interesting to note that such a multi-step
transition behavior is not limited to the quarter-filled system. At
a more general filling fraction $\frac{1}{qp}$, the system can first
enter a $\frac{1}{p}$($\frac{1}{q}$)-filled intermediate metallic
phase by a $q$($p$)-times enlargement of unit cell and then arrive
at the integer-filled true insulating state by a second transition
with a $p$($q$)-times enlargement of the unit cell.

\begin{figure}[h!]
\caption{The finite
temperature phase diagram of the model for both signs of the hopping
integral.(a)$t>0$, (b)$t<0$. } \label{fig3}
\end{figure}

\begin{figure}[h!]
\caption{The jump of the order parameter at the
collinear-non-coplanar transition point as a function of $U/|t|$ for
both signs of the hopping integral. } \label{fig4}
\end{figure}

In the intermediate state between 53K and 88K, the collinear spin
ordering induces dimensional reduction in the charge transport. In
fact, all metal-insulator transition can be viewed as a dimensional
reduction process in the charge sector: an insulator can be viewed
as a system in which the charge transport is quenched in all
dimension. Such a dimensional reduction process in the charge sector
can occur either directly, or like our example, in a multi-step
manner in spatial dimension larger than 2. In the intermediate state
, the charge transport is quenched in some dimension but remain
active in other dimension. Such dimensional reduction in the bulk of
the system can occur in condensed matter system with multiple low
energy degree of freedoms. One famous example is the orbital
ordering induced dimensional reduction in ruthenium oxides in which
the electron hopping is blocked except in one direction as a result
of the symmetry property of the electron orbital wave function.
Another example is the stripe-like structure in some cuprates
superconductors in which the charge motion becomes effectively one
dimensional as a result of microscopic phase separation\cite{nco13}.

Finally, we discuss the relevance of our results to the physics of
the $Na_{0.5}CoO_{2}$ system. The quasi-one dimensional metallic
state with collinear magnetic order has the proper characteristics
to account for the experimental observations between 53K and 88K.
Thus it is not unreasonable to assign this state to the intermediate
temperature region. However, as our model is purely two-dimensional,
additional three dimensional coupling should be invoked to explain
the finite temperature existence of the magnetic order in the
experiment. Here we assume that such three dimensional coupling will
not alter the essential physics discussed in this paper.
Furthermore, as our treatment of the model is limited to the mean
field level, it is important to know how fluctuation effect will
change the results. We leave this important issue to future study.
As to the insulating state below 53K, we note that in real system,
the perfect hexagonal symmetry maybe broken by perturbation such as
the sodium order potential, or spontaneously through the Peries
instability process of the quasi-one dimensional intermediate state,
the real system have many different ways to become a true insulator.
It is important to make sure that the non-coplanar state studied in
this paper is stable with respect to these perturbations before
assigns it as the true ground state of the $Na_{0.5}CoO_{2}$ system.

T. Li is supported by NSFC Grant No. 10774187, National Basic
Research Program of China No. 2007CB925001 and Beijing Talent
Program.

\bigskip


\begin{thebibliography}{99}
\bibitem{qunit} $\vec{\mathbf{b}}_{1}=\frac{4\pi}{\sqrt{3}a}(0,1)$ and $\vec{\mathbf{b}}_{2}=\frac{2\pi}{a}(1,-\frac{1}{\sqrt{3}})$ are the two reciprocal lattice vectors of the triangular lattice.
\end{thebibliography}
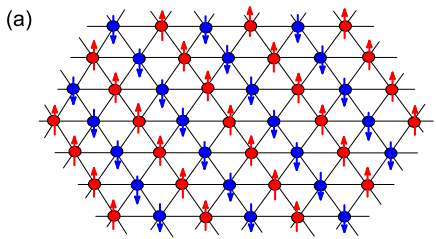
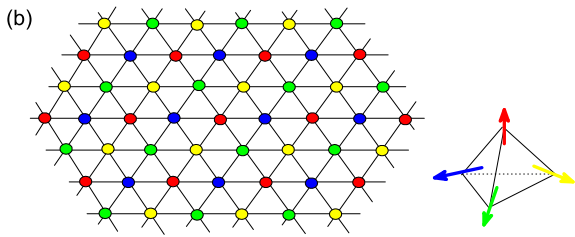
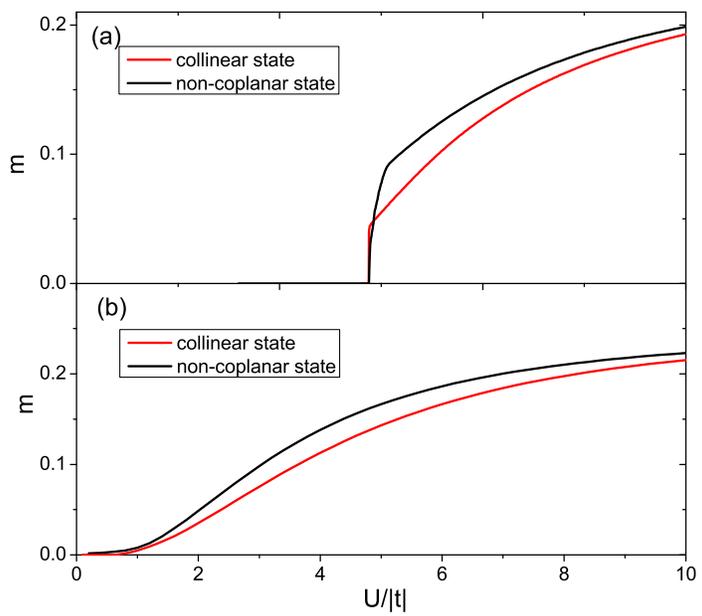
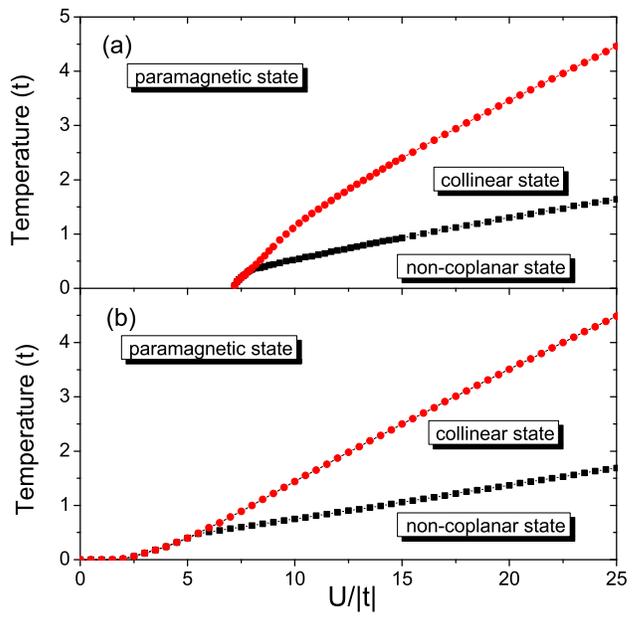
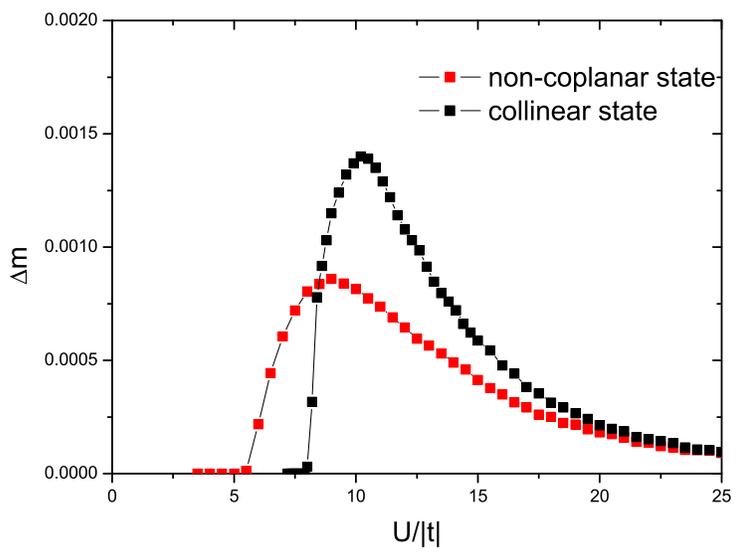
\end{document}